\documentclass[a4paper]{jpconf}
\usepackage{graphicx}

\begin{document}
\title{Main features of detectors and isotopes to investigate double beta decay with increased sensitivity}

\author{A S Barabash}
\address{National Research Center "Kurchatov Institute", Institute of Theoretical and Experimental Physics, B. Cheremushkinskaya 25, 117218 Moscow, Russia}


\begin{abstract}
The current situation in double beta decay experiments , the characteristics of modern detectors and the possibility of increasing the sensitivity to neutrino mass in future experiments are discussed. The issue of the production and use of enriched isotopes in double beta decay experiments is discussed in addition.
\end{abstract}

\section{Introduction}
Interest in neutrinoless double-beta decay has seen a significant renewal after 
evidence for neutrino oscillations was obtained from the 
results of atmospheric, solar, reactor and accelerator  neutrino 
experiments (see, for example, discussions in \cite{MOH06,BIL15,VAL17}). 
These results are impressive proof that neutrinos have a non-zero mass. However,
the experiments studying neutrino oscillations are not sensitive to the nature
of the neutrino mass (Dirac or Majorana) and provide no information on the 
absolute scale of the neutrino masses, since such experiments are sensitive 
only to the difference of the masses, $\Delta m^2$. The detection and study 
of $0\nu\beta\beta$ decay may clarify the following problems of neutrino 
physics (see discussions in \cite{PAS06,BIL15a,SIM16}):
 (i) lepton number non-conservation, (ii) neutrino nature: 
whether the neutrino is a Dirac or a Majorana particle, (iii) absolute neutrino
 mass scale, (iv) the type of neutrino 
mass hierarchy (normal, inverted, or quasidegenerate), (v) CP violation in 
the lepton sector (measurement of the Majorana CP-violating phases).

Let us consider three main modes of $\beta\beta$ decay:

\begin{equation}
(A,Z) \rightarrow (A,Z+2) + 2e^{-} + 2\bar { \nu},
\end{equation}

\begin{equation}
(A,Z) \rightarrow (A,Z+2) + 2e^{-},
\end{equation}

\begin{equation}
(A,Z) \rightarrow (A,Z+2) + 2e^{-} + \chi^{0}(+ \chi^{0}).
\end{equation}

\underline{The $2\nu\beta\beta$ decay (process (1))} is a second-order
process, which is not forbidden
by any conservation law. The half-life can be written as 
\begin{equation}
   [T_{1/2}(2\nu)]^{-1} = G_{2\nu} g_A^4 \mid{M_{2\nu}}\mid^2 ,
\end{equation}
where $G_{2\nu}$ is the phase space factor (which is accurately known \cite{KOT12,STO15}), $\mid{M_{2\nu}}\mid^2$ is the nuclear matrix element (NME) 
and $g_A$ is the axial-vector coupling constant\footnote{Usually the value $g_A$ = 1.27 is used (the free neutron decay value). In nuclear matter, however, the value
of $g_A$ could be quenched. In $2\nu$ decay case effect of quenching could be quite strong \cite{SIM16,BAR15,PIR15,KOS17,SUH17}. In case of $0\nu$ decay it can be a factor $\sim$ 1.2-1.5 (see discussion in \cite{SUH17}). This question is still under discussion and there is no final answer up to now.}. 
The detection of this process provides the
experimental determination  of the 
NMEs involved
in the double beta decay  processes.  This leads to the development of
theoretical schemes for NME calculations  both in
connection with the $2\nu\beta\beta$ decays as well as the
$0\nu\beta\beta$ decays \cite{ROD06,ROD06a,KOR07,KOR07a,SIM08,BAR13}. Precise measurement of 
$2\nu\beta\beta$ decay rate for different nuclei can help to solve $g_A$ problem 
(see discussions in \cite{SIM16,BAR15,PIR15,KOS17,SUH17}).  
Moreover, the study could yield an investigation of the
time dependence of the weak interaction coupling constant 
\cite{BAR98,BAR00,BAR03}. 

Recently, it has been pointed out that the $2\nu\beta\beta$ decay allows 
one to investigate particle properties, in particular whether the Pauli 
exclusion principle is violated for neutrinos and thus neutrinos partially obey
Bose--Einstein statistics \cite{DOL05,BAR07b}. In addition Lorentz and 
CPT violation could be checked in double beta decay by investigation 
of 2$\nu$ spectrum modification (see, for example, \cite{ALB16}).

\underline{The $0\nu\beta\beta$ decay (process (2))} violates the law
of lepton-number conservation
($\Delta L =2$)
and requires that the Majorana neutrino has a nonzero mass. Also, this
process is possible in some
supersymmetric models, where $0\nu\beta\beta$ decay is initiated by
the exchange of supersymmetric
particles. This decay also arises in models featuring an extended
Higgs sector within
electroweak-interaction theory and in some other cases
(see, for example, \cite{KLA98,SIM16}). In the case of neutrino mass mechanism 
the half-life can be written as
\begin{equation}
   [T_{1/2}(0\nu)]^{-1} = G_{0\nu}g_A^4
   \mid{M_{0\nu}}\mid^2\Bigl|{\frac{\langle m_{\nu}\rangle}{m_e}}\Bigr|^2,
\end{equation}
where $G_{0\nu}$ is the phase space factor, which contains 
the kinematic information about the final state particles, 
and is exactly calculable to the precision of the input 
parameters \cite{KOT12,STO15}, $g_A$ is the axial-vector coupling constant,
$\mid{M_{0\nu}}\mid^2$ is the nuclear 
matrix element, $m_e$ is the mass of the electron, and 
$\langle m_{\nu}\rangle$ is the effective Majorana mass 
of the electron neutrino, which is defined as 
$\langle m_{\nu}\rangle$ = $\mid\sum_iU^2_{ei}m_i\mid$ where $m_i$ 
are the neutrino mass eigenstates and $U_{ei}$ 
are the elements of the neutrino mixing 
Pontecorvo-Maki-Nakagawa-Sakata (PMNS) matrix.

\underline{The $0\nu\chi^{0}\beta\beta$ decay (process (3))} requires
the existence of a Majoron. It is a massless
Goldstone boson that arises due to a global breakdown of ($\it{B}$-$\it{L}$)
symmetry, where {\it B} and {\it L} are,
respectively, the baryon and the lepton number. The Majoron, if it
exists, could play a significant role
in the history of the early Universe and in the evolution of
stars. In addition, Majoron could play the role of the dark matter particle (see, for example, \cite{LAT13}). The model of a triplet
Majoron \cite{GEL81} was disproved in 1989
by the data on the decay width of the $Z^{0}$ boson that were
obtained at the LEP accelerator \cite{CAS98}. Despite this, some new models were proposed
\cite{MOH91,BER92}, where $0\nu\chi^{0}\beta\beta$
decay is possible
and where there are no contradictions with the LEP data. A
$\beta\beta$-decay model that involves the
emission of two Majorons was proposed within supersymmetric
theories \cite{MOH88} and several other models of the
Majoron were proposed in the 1990s. By the term ``Majoron'', one
means massless or light bosons
that are associated with neutrinos. In these models, the Majoron
can carry a lepton charge and is
not required to be a Goldstone boson \cite{BUR93,BUR94}. A decay process
that involves the emission of two ``Majoron''
is also possible \cite{BAM95}. In models featuring a vector
Majoron, the Majoron is the longitudinal
component of a massive gauge boson emitted in $\beta\beta$ decay
\cite{CAR93}. For the sake of simplicity, each such
object is referred to here as a Majoron.
In the Ref. \cite{MOH00}, a ``bulk'' Majoron model was proposed
in the context of the ``brane-bulk'' scenario for particle physics.

\begin{figure}[b]
\centerline{\includegraphics[width=7cm]{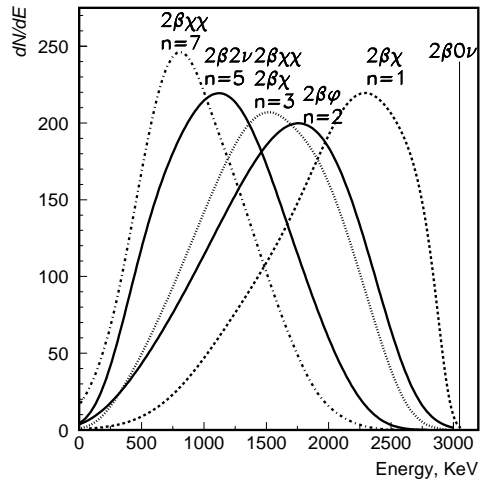}}
\caption{Energy spectra of different modes: $2\nu\beta\beta$
$(n=5)$,
$0\nu\chi^{0}\beta\beta$ $(n=1~$,$~2$ and$~3)$ and
$0\nu\chi^{0}\chi^{0}\beta\beta (n=3~$and$~7)$ decays of $~^{100}$Mo.\label{f1}}
\end{figure}

The possible sum two electrons energy spectra for different $\beta\beta$
decay modes of $^{100}$Mo are shown
in Fig.~\ref{f1}. Here {\it n} is the spectral 
index, which defines
the shape of the spectrum. For example, for an ordinary Majoron {\it n} = 1, 
for 2$\nu\beta\beta$ decay {\it n} = 5, in the case of a bulk Majoron {\it n} = 2 
and for the process with two Majoron emission {\it n} = 3 or 7.
The half-life for ordinary Majoron with spectral index {\it n} = 1 can be written as 
\begin{equation}
   [T_{1/2}(0\nu\chi^0)]^{-1} = G_{0\nu\chi^0} g_A^4\langle g_{ee}\rangle^2 \mid{M_{0\nu}}\mid^2, 
\end{equation}
where $G_{0\nu\chi^0}$ is the phase space factor (which is accurately known \cite{KOT15,KOT15a}), $\mid{M_{0\nu}}\mid^2$ is the nuclear matrix element (the same as for $0\nu\beta\beta$ decay), 
$\langle g_{ee}\rangle$ is the coupling constant of the Majoron to the neutrino
and $g_A$ is the axial-vector coupling constant. 

\section{Results of experimental investigations}

\begin{figure}[b]
\centerline{\includegraphics[width=7cm]{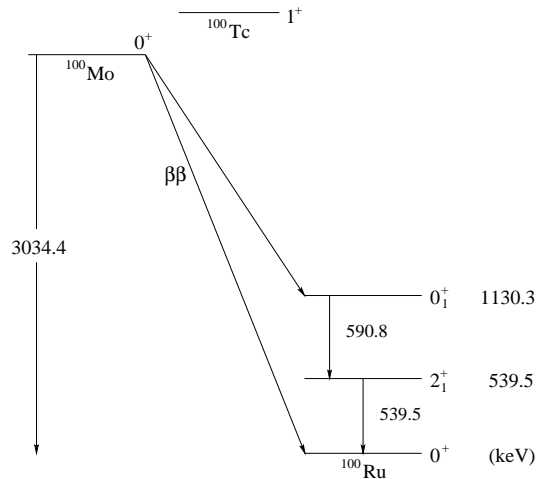}}
\caption{Simplified levels scheme for $^{100}$Mo-$^{100}$Tc-$^{100}$Ru.\label{f2}}
\end{figure}

The number of possible candidates for double beta decay is quite
large, there are 35 nuclei\footnote{In addition 34
nuclei can undergo double electron
capture, while twenty two nuclei and six nuclei can undergo,
respectively, EC$\beta^{+}$ and 2$\beta^{+}$ decay (see the tables
in \cite{TRE02}).}. However, nuclei for which the double beta transition energy
($E_{\beta\beta}$) is in excess
of 2 MeV are of greatest interest, since the double beta decay
probability strongly depends on the transition
energy ($\sim E^{11}_{\beta\beta}$ for $2\nu\beta\beta$ decay, 
$\sim E^{7}_{\beta\beta}$ for $0\nu\chi^{0}\beta\beta$ decay and 
$\sim E^{5}_{\beta\beta}$ for $0\nu\beta\beta$ decay). 
In transitions to excited states of the daughter nucleus,
the excitation energy is removed via the
emission of one or more photons, which can be detected, and this
can serve as an additional source
of information about double beta decay. As an example Fig.~\ref{f2}
shows the diagram of energy levels in the
$^{100}$Mo-$^{100}$Tc-$^{100}$Ru nuclear triplet.

\subsection{Two neutrino double beta decay}

This decay was first recorded in 1950 in a geochemical experiment
with $^{130}$Te \cite{ING49}. In 1967,
$2\nu\beta\beta$ decay was found for $^{82}$Se also in a geochemical
experiment \cite{KIR67}.
Attempts to observe this decay in a direct experiment employing
counters had been futile for
a long time. And only in 1987, M. Moe and his co-authors observed for the first time the $2\nu\beta\beta$  decay of $^{82}$Se using a time-projection chamber (TPC) \cite{ELL87}. 
In the next few years,
experiments were able to detect
$2\nu\beta\beta$
decay in many nuclei. In $^{100}$Mo and $^{150}$Nd 
$2\beta(2\nu)$ decay to the $0^{+}$ excited state of the
daughter nucleus was measured too (see \cite{BAR15a}). Also, the
$2\nu\beta\beta$ decay of $^{238}$U was detected in a radiochemical
experiment \cite{TUR91}, and in geochemical experiments
the ECEC process was detected in $^{130}$Ba \cite{MES01,PUI09}. 
Table 1 displays the present-day averaged and
recommended $T_{1/2}$(2$\nu$) values from \cite{BAR15a}.
At present, experiments devoted to detecting
$2\nu\beta\beta$ decay are approaching a level
where it is insufficient just to record the
decay. It is necessary to
measure several parameters of this process to a high precision 
(half-life value, energy sum spectrum, single electron energy spectrum and angular 
distribution).
Tracking detectors that are able to record both
the energy of each electron and the angle at which they diverge
are the most appropriate instruments for this.

\begin{table}[ht]
\label{Table1}
\caption{Average and recommended $T_{1/2}(2\nu)$ values (from
\cite{BAR15a}).}
\begin{center}
\begin{tabular}{ll} 
\hline
Isotope & $T_{1/2}(2\nu)$, yr \\ 
\hline
$^{48}$Ca & $4.4^{+0.6}_{-0.5}\times10^{19}$ \\
$^{76}$Ge & $1.65^{+0.14}_{-0.12}\times10^{21}$ \\
$^{82}$Se & $(0.92 \pm 0.07)\times10^{20}$ \\
$^{96}$Zr & $(2.3 \pm 0.2)\times10^{19}$ \\
$^{100}$Mo & $(7.1 \pm 0.4)\times10^{18}$ \\
$^{100}$Mo-$^{100}$Ru$(0^{+}_{1})$ & $6.7^{+0.5}_{-0.4}\times10^{20}$ \\
$^{116}$Cd & $(2.87 \pm 0.13)\times10^{19}$\\
$^{128}$Te & $(2.0 \pm 0.3)\times10^{24}$ \\
$^{130}$Te & $(6.9 \pm 1.3)\times10^{20}$ \\
$^{136}$Xe & $(2.19 \pm 0.06)\times10^{21}$ \\
$^{150}$Nd & $(8.2 \pm 0.9)\times10^{18}$ \\
$^{150}$Nd-$^{150}$Sm$(0^{+}_{1})$ & $1.2^{+0.3}_{-0.2}\times10^{20}$
\\
$^{238}$U & $(2.0 \pm 0.6)\times10^{21}$  \\
$^{130}$Ba; ECEC(2$\nu$) & $\sim 10^{21}$  \\ 
\hline

\end{tabular} 
\end{center}
\end{table}

\subsection{Neutrinoless double beta decay}

In contrast to two-neutrino decay, neutrinoless double beta decay
has not yet been observed, although it is easier (from experimental point of view)
to detect it. In this case, one seeks, in the
experimental spectrum, a peak of energy equal to the double beta
transition energy and of width determined
by the detector's resolution.

The constraints on the existence of
$0\nu\beta\beta$ decay are presented in Table 2 for the nuclei
that are the most promising candidates. Limits on $\langle m_{\nu} \rangle$ are given using recent (most reliable) NME calculations and $g_A$ = 1.27. The spread of the $\langle m_{\nu} \rangle$ values is associated with uncertainties in the NME calculations.

The best limits on 
$\langle m_{\nu}\rangle$ are obtained for $^{136}$Xe, $^{76}$Ge, $^{130}$Te and $^{100}$Mo. 
The assemblage of sensitive experiments
for different nuclei permits one to increase the reliability of the limit 
on $\langle m_{\nu}\rangle$. Present conservative limit can be set as 0.24 eV 
(using conservative value from the KamLAND-Zen experiment). But one has to take into account that, in fact, this value could be in $\sim$ 1.5-2 times greater because of the possible quenching of $g_A$ (see recent discussions in \cite{SUH17}).

\begin{table}[ht]
\label{Table2}
\caption{Best present limits on $0\nu\beta\beta$ decay (at
90\% C.L.). To calculate $\langle m_{\nu}\rangle$ 
the NME from \cite{SUH15,SIM13,BAR15,RAT10,ROD10,MEN09,HOR15,MUS13,SON17}, 
phase-space factors from \cite{KOT12,STO15} and $g_A$ = 1.27 have been used.
In case of $^{150}$Nd NME from \cite{TER15,FAN15} 
and in case of $^{48}$Ca from \cite{IWA16} were used in addition. }
\hspace{0.2cm}
\begin{center}
\begin{tabular}{lllll}
\hline
Isotope & Q$_{2\beta}$, keV & $T_{1/2}$, yr & $\langle m_{\nu} \rangle$, eV 
 & Experiment \\ 
\hline
$^{48}$Ca & 4267.98  & $>5.8\times10^{22}$ & $<3.1-15.4$ & CANDLES ~\cite{UME08} \\
$^{76}$Ge & 2039.00 & ${\bf >5.1\times10^{25}}$ & ${\bf <0.15-0.39}$ & GERDA-I+GERDA-II ~\cite{AGO17a} \\
& & ($>8\times10^{25}$) & ($<0.12-0.31$) & \\
$^{82}$Se &2997.9 & $>3.6\times10^{23}$ & $<1-2.4$ & NEMO-
3 ~\cite{BAR11} \\
$^{96}$Zr & 3355.85 & $>9.2\times10^{21}$ & $<3.6-10.4$ & NEMO-3 ~\cite{ARN10} \\
$^{100}$Mo & 3034.40 & $>1.1\times10^{24}$ & $<0.33-0.62$ & NEMO-
3 ~\cite{ARN15} \\
$^{116}$Cd & 2813.50 & $>2.4\times10^{23}$ & $<0.9-1.6$ & AURORA ~\cite{POL17} \\
$^{128}$Te & 866.6 & $>1.5\times10^{24}$ & $2.3-4.6$ & Geochem. exp. (see ~\cite{BAR15a})  \\
$^{130}$Te & 2527.52 & ${\bf >0.8\times10^{25}}$ & ${\bf <0.18-0.69}$ & CUORICINO + CUORE0  \\
&  & ($>1.5\times10^{25}$) & ($<0.13-0.50$) & + CUORE ~\cite{ALD17} \\
$^{136}$Xe & 2457.83 & ${\bf >0.5\times10^{26}}$ & ${\bf <0.09-0.24}$ & KamLAND-Zen
~\cite{GAN16} \\
 & & ($>1.07\times10^{26}$) & ($<0.06-0.16$) & \\
$^{150}$Nd & 3371.38 & $>2\times10^{22}$ & $<1.6-5.3$ & NEMO-3 ~\cite{ARN16} \\ 
\hline
\end{tabular} 
\end{center}
\end{table}

The main goal of next generation experiments is to investigate inverted hierarchy region of neutrino mass ($\langle m_{\nu} \rangle$ $\approx$ (15-50) meV). If one will not see the decay in this region then it will be necessary to investigate region with $\langle m_{\nu} \rangle$  $<$ 15 meV (normal hierarchy region).

\subsection{Double beta decay with Majoron emission}

Table 3 displays the best present-day constraints for an ``ordinary''
Majoron ({\it n} = 1).
Table 4 gives the best experimental
constraints on decays accompanied by the emission of one or two
Majorons (for {\it n} = 2, 3, and 7).
Hence at the present time only limits on double beta decay with
Majoron emission have been obtained (see Table 3 and Table 4).
A conservative present limit on the coupling constant of ordinary 
Majoron to the neutrino is $\langle g_{ee} \rangle < 1.2 \times 10^{-5}$.
But again, one has to take into account that, in fact, this value could be in $\sim$ 1.5-2 times greater because of the possible quenching of $g_A$ (see recent discussions in \cite{SUH17}).

\begin{table}[ht]
\label{Table3}
\caption{Best present limits on $0\nu\chi^{0}\beta\beta$ decay
(ordinary Majoron; {\it n} = 1) at 90\% C.L. To calculate $\langle g_{ee} \rangle$ 
the NME from \cite{SUH15,SIM13,BAR15,RAT10,ROD10,MEN09,HOR15,MUS13,SON17}, 
phase-space factors from \cite{KOT15,KOT15a} and $g_A$ = 1.27 have been used. In case of $^{150}$Nd NME from \cite{TER15,FAN15} 
and in case of $^{48}$Ca from \cite{IWA16} were used in addition.}
\hspace{0.2cm}
\begin{center}
\begin{tabular}{llll} 
\hline
Isotope & $T_{1/2}$, yr & $\langle g_{ee} \rangle$, $\times 10^{-5}$
 & Experiment \\  
\hline
$^{48}$Ca & $>4.6\times10^{21}$  & $< 8.6-43.1 $ & NEMO-3 ~\cite{ARN16b} \\
$^{76}$Ge & $>4.2\times10^{23}$  & $< 2.4-6.3 $ & GERDA-I ~\cite{AGO15} \\
$^{82}$Se & $>1.5\times10^{22}$  & $< 5.0-12.2 $ & NEMO-3 ~\cite{ARN06}\\
$^{96}$Zr & $>1.9\times10^{21}$  & $7.4-21.3$  & NEMO-3 ~\cite{ARN10} \\
$^{100}$Mo & $>4.4\times10^{22}$ & $< 1.7-3.1 $ & NEMO-3 ~\cite{ARN15} \\
$^{116}$Cd & $>1.1\times10^{22}$  & $4.5-8.0$  & AURORA ~\cite{POL17} \\
$^{128}$Te & $>1.5\cdot10^{24}$  & $6.3-12.3$ & Geochem. exp. (see ~\cite{BAR15a})  \\
$^{130}$Te & $>1.6\times10^{22}$  & $< 4.6-17.4 $ & NEMO-3 ~\cite{ARN11} \\
$^{136}$Xe & $> 2.6\times10^{24}$ & $< 0.45-1.2 $ & KamLAND-Zen ~\cite{GAN12} \\
$^{150}$Nd & $>3\times10^{21}$  & $< 3.7-11.9 $ & NEMO-3 ~\cite{ARN16} \\ 
\hline
\end{tabular}
\end{center}
\end{table}

\begin{table}[ht]
\label{Table4}
\caption{Best present limits on $T_{1/2}$ (yr) for decay with one and
two Majorons at 90\% C.L. for modes
with spectral index {\it n} = 2, {\it n} = 3 and {\it n} = 7.}
\hspace{0.2cm}
\begin{center}
\begin{tabular}{llll} 
\hline
Isotope ($E_{2\beta}$, keV) & {\it n} = 2 & {\it n} = 3 &  {\it n} = 7  \\ 
\hline
$^{76}$Ge (2039) & $>1.8\times10^{23}$ ~\cite{AGO15} & $>8\times10^{22}$ ~\cite{AGO15} &
$>3\times10^{22}$ ~\cite{AGO15} \\
$^{82}$Se (2998) & $>6\times10^{21}$ ~\cite{ARN06} & $>3.1\times10^{21}$
~\cite{ARN06} & $>5\times10^{20}$ ~\cite{ARN06} \\
$^{96}$Zr (3356) & $>9.9\times10^{20}$ ~\cite{ARN10} & $>5.8\times10^{20}$ ~\cite{ARN10} &
$>1.1\times10^{20}$ ~\cite{ARN10} \\
$^{100}$Mo (3034) & $>1.7\times10^{22}$ ~\cite{ARN06} & $>1\times10^{22}$
~\cite{ARN06} & $>7\times10^{19}$ ~\cite{ARN06} \\
$^{116}$Cd (2813) & $>2.1\times10^{21}$ ~\cite{POL17} & $>0.9\times10^{21}$
~\cite{POL17} & $>4.1\times10^{19}$ ~\cite{ARN06} \\
$^{128}$Te (867) & $>1.5\times10^{24}$ ~\cite{MAN91,BAR15a}  & $>1.5\times10^{24}$ ~\cite{MAN91,BAR15a} 
& $>1.5\times10^{24}$ ~\cite{MAN91,BAR15a} \\
(geochem)  &  &   & \\
$^{130}$Te (2528) & - & $>9\times10^{20}$ ~\cite{ARN03} & - \\
$^{136}$Xe (2458) & $>1.0\times10^{24}$ ~\cite{GAN12} & $>4.5\times10^{23}$ ~\cite{GAN12} & 
$>1.1\times10^{22}$ ~\cite{GAN12} \\
$^{150}$Nd (3371) & $>5.4\times10^{20}$ ~\cite{ARG09} & $>2.2\times10^{20}$ ~\cite{ARG09} &
$>4.7\times10^{19}$ ~\cite{ARG09} \\ 
\hline
\end{tabular}
\end{center}
\end{table}

\section{Main features of detectors to investigate neutrinoless double beta decay with increased sensitivity}

\subsection{Requirements to the detectors}
In case of no signal, the half-life sensitivity for a $0\nu\beta\beta$ experiment is given by:
$$
T_{1/2}(0\nu) > \frac{0.69\times N\times m\times \eta\times t}{n_{exc}}
$$
where {\it m} is the molar mass of enriched isotope, {\it N} is the Avogadro number, $\eta$ is the $0\nu\beta\beta$ efficiency of the detector, {\it t} is the duration of the measurement 
and {\it n$_{exc}$} is the number
of excluded $0\nu\beta\beta$ events (which start to be higher with increasing of background). From this relation, it is clear that 
the essential requirements for the
double beta experiments are to achieve an extremely low radioactive background, a large mass
of isotope, a large detection efficiency and a long measurement time. It is also important to note that in the case of no
background, the half-life sensitivity increases as the duration
of observation t , while in the
case of relatively larger background, it increases only as $\surd t$.

The basic requirements to the next-generation detectors for searching for  $0\nu\beta\beta$ decay can be formulated as follows:

1) The large mass of the investigated isotope. In the next generation experiments investigating the region of the inverse hierarchy, the mass of the isotope should be on the level of 100-1000 kg. To study the area of normal hierarchy, weights of 10-100 tons could be required (see section 5).

2) The background in the region of interest (ROI) should be minimal. The ideal case is when the background is zero. To achieve a low background level, the detector must be located deep underground, surrounded by passive and active shields. The construction materials used must contain a minimum amount of radioactive isotopes. The background index (BI) in the best modern experiments is $\sim$ 10$^{-3}$ counts/keV$\times$kg$\times$yr. In planned next-generation experiments, it is proposed to reduce the background up to $\sim 10^{-4}-10^{-5}$ c/keV$\times$kg$\times$yr, and to search for  $0\nu\beta\beta$ decay in the normal hierarchy region up to $\sim 10^{-6}$ c/keV$\times$kg$\times$yr. Here it should be noted that in one modern experiment (KamLAND-Zen) the background index on the level of $\sim 10^{-5}$ c/keV$\times$kg$\times$yr has already been achieved. Of course, KamLAND-Zen is a very specific case, but it shows that such a background level is quite achievable.

3) It is desirable to have detector characteristics that help to reduce the background level in the ROI, to distinguish the useful events from the background, to measure the various parameters of the double beta decay process and to ensure the detector's uninterrupted operation for a long time (5-10 years). One of the key characteristics is the energy resolution. Good energy resolution can significantly reduce the size of ROI and thereby reduce the number of background events in the ROI. A good energy resolution also allows to minimize the ineradicable background from $2\nu\beta\beta$  decay. At an energy resolution of better than (1-2)\%, the background from $2\nu\beta\beta$  decay (in all practical cases) can be neglected. Very useful are such properties of the detector as spatial resolution, granularity, registration of the pulse shape of the signal, detection of electron tracks, etc.

Table 5 lists the main current experiments and most developed and promising projects.

\begin{table}[ht]
\caption{Main current experiments and most developed and promising projects. 
Sensitivity at 90\% C.L. for three (GERDA-II, Majorana Demonstrator, 
first step of SuperNEMO and KamLAND-Zen) and
five (EXO-200, nEXO, SuperNEMO, SNO+, KamLAND-Zen and CUORE)  
years of measurements is presented. M is mass of the isotope.}
\hspace{0.2cm}
\begin{center}
\begin{tabular}{llllll} 
\hline
Experiment & Isotope & M, kg & Sensitivity & Sensitivity & Status \\
& &  & $T_{1/2}$, yr & $\langle m_{\nu} \rangle$, meV &  \\ 
\hline
CUORE ~\cite{ART15}  & $^{130}$Te & 200 & $9.5\times10^{25}$ & 53--200 & current \\  
GERDA-II ~\cite{AGO17} & $^{76}$Ge & 35 & $1.5\times10^{26}$ & 90--230 & current \\ 
Majorana-D ~\cite{ABG14}  & $^{76}$Ge & 30 & $1.5\times10^{26}$ & 90--230 & current \\
LEGEND ~\cite{ABG17b} & $^{76}$Ge & 200 & $\sim 10^{27}$ & 34--90 & R\&D \\
       &  & 1000 & $\sim 10^{28}$ & 11-28 & R\&D \\
EXO ~\cite{ALB14,ALB17} & $^{136}$Xe & 200 & $5.7\times10^{25}$ & 85--225 & current \\
& & 5000 & $6\times 10^{27}$ & 8--22 & R\&D \\ 
SuperNEMO ~\cite{BAR17,CAS16} & $^{82}$Se & 7 & 6.5$\times10^{24}$ & 240--560 & in progress \\
  &  & 100--140 & $\sim (1-1.5)\times10^{26}$ & 50--140 & R\&D \\
KamLAND-Zen ~\cite{SHI17}  & $^{136}$Xe & 750 & 2$\times10^{26}$ & 45--120 & in progress \\
 & & 1000 &  $6\times10^{26}$ & 25-70 & R\&D \\
SNO+ ~\cite{AND16,ARU17} & $^{130}$Te & 1300 & $1.9\times10^{26}$ & 40--140 & in progress \\
 & & 8000 & $\sim 7\times 10^{26}$ & 20-70 &  R\&D \\
\hline
\end{tabular}
\end{center}
\end{table}

\subsection{Types of the detectors}

\subsubsection{Germanium detectors}
The use of semiconductor germanium detectors to search for neutrinoless double beta decay has a long and rich history. The main reasons for such popularity of the germanium detectors are availability (they are manufactured by industry), almost 100\% efficiency of registration of $\beta\beta$ events from $^{76}Ge$ and absolutely unique energy resolution of such detectors ($\sim$ 0.1\% FWHM at 2 MeV). In addition, the level of radioactive impurities in germanium crystals is extremely low (this is obtained "automatically" during initial germanium purification and crystal growth). For the first time, the Ge(Li) detector was used to search for the double beta decay of $^{76}$Ge in 1967 by E. Fiorini, and the limit $T_{1/2} (0\nu) > 3\times10^{20}$ yr was obtained \cite{FIO67}. In 1973, the limit was improved up to $T_{1/2}(0\nu) > 5\times10^{21}$ years 
\cite{FIO73}. In 1987, semiconductor germanium detectors grown from enriched germanium (still Ge(Li) detectors) were first used, which led to a sharp increase in the sensitivity of measurements \cite{VAS88,VAS90}. Since the beginning of the 1980s, High Purity Germanium (HPGe) detectors usually have been used, which are much more convenient to operate. As a result, in the 1990s, two independent experiments were carried out using HPGe detectors made of enriched germanium ($^{76}$Ge content is 86\%) - the Heidelberg-Moscow experiment (10 kg of enriched germanium) and IGEX (6.5 kg of enriched germanium). The best (by that time) limits were obtained: $T_{1/2}(0\nu) > 1.9\times10^{25}$ yr \cite{KLA01} and $T_{1/2}(0\nu)> 1.57\times10^{25}$ yr \cite{AAL02}, respectively. The natural continuation of this activity was the GERDA and Majorana experiments.

\underline{The GERDA experiment.} The GERDA experiment is located at the underground Laboratori Nazionali del Gran Sasso (LNGS) of INFN, Italy. A rock overburden of about 3500 m w. e. removes the hadronic components of cosmic ray showers and reduces the muon flux at the experiment by six orders of magnitude. The basic idea is to operate bare germanium detectors in a radiopure cryogenic liquid like LAr for cooling to their operating temperature of $\sim$ 90 K and for shielding against external radiation originating from the walls. In GERDA, a 64 m$^3$ LAr cryostat is inside a 590 m$^3$ water tank. The clean water completes the passive shield. Above the water tank is a clean room with a glove box and lock for the assembly of germanium detectors into strings and the integration of the liquid argon veto system. 


A first physics data taking campaign (Phase I) was carried out from November 2011 to June 2013 using mainly coaxial type detectors (previously used in Heidelberg-Moscow and IGEX experiments)  and the results showed no indication of a $0\nu\beta\beta$ signal \cite{AGO13}. After the completion of Phase I the GERDA setup has been upgraded to perform its next step (Phase II): the goal was the tenfold reduction of the background, optimizing the experimental setup, and the increase of the $^{enr}$Ge detector mass. Thirty Broad Energy Germanium (BEGe) detectors from Canberra were deployed: they allow superior background rejection and have an excellent energy resolution. In addition an active suppression of background by detecting the LAr scintillation light, consisting of PMTs and wavelength shifting fibers coupled to silicon photomultipliers, has been introduced. On December 20th, 2015 the Phase II data taking with all 40 detectors (30 BEGes, 7 $^{enr}$Ge and 3 $^{nat}$Ge coaxial detectors) started. The first Phase II data were released after 6 months, corresponding to an exposure of 10.8 kg$\times$yr of $^{enr}$Ge (5.0 kg$\times$yr from $^{enr}$Ge coaxial detectors and 5.8 kg$\times$yr from BEGes) \cite{AGO17}.   In 2017 new result has been published. The exposure available for analysis was 23.5 kg$\times$yr and 23.2 kg$\times$yr for Phase I and Phase II, respectively, totaling (236.9 $\pm$ 4.6) mol$\times$yr of $^{76}$Ge in the active volume of the detectors. Using this data, a new limit $T_{1/2} > 8\times10^{25}$ yr at 90\% C.L. (or $T_{1/2} > 5.1\times10^{25}$ yr at 90\% C.L. within the Bayesian framework) \cite{AGO17a} has been obtained. The average energy resolution at Q$_{\beta\beta}$ for BEGe and coaxial detectors was 2.9 keV and 3.9 keV, respectively, background index was $\sim 10^{-3}$ c/keV$\times$kg$\times$yr. 

GERDA is currently accumulating data. Taking into account the background level achieved
so far and 11.2 kg$\times$yr of still-to-be-unblinded coaxial data, the median sensitivity will break the
wall of $10^{26}$ yr within 2018. The sensitivity expected for the full 100 kg$\times$yr exposure of GERDA is $\sim 1.5\times10^{26}$ yr.

\underline{The Majorana Demonstrator experiment.} The Majorana Demonstrator is an array of isotopically enriched and natural Ge detectors 
that will search for the double beta decay of $^{76}$Ge \cite{ABG14}. The primary goal of the Demonstrator is to demonstrate a path forward to achieving a background rate low enough (at or below 1 count/(ROI$\times$t$\times$yr) in the 4 keV region of interest (ROI) around the 2039-keV (Q$_{\beta\beta}$-value for $^{76}$Ge) to ensure the feasibility of a future Ge-based $\beta\beta$ experiment to probe the inverted-hierarchy parameter space of neutrino mass. 
Majorana utilizes the demonstrated benefits of enriched HPGe 
detectors. These include intrinsically low-background source material, understood enrichment chemistry, excellent energy resolution, and sophisticated event reconstruction. It was 
assembled a modular instrument composed of two cryostats built from ultra-pure electroformed 
copper, with each cryostat housing over 20 kg of P-type, point-contact (P-PC) HPGe 
detectors. P-PC detectors were chosen after extensive R\&D and each has a mass of about 0.6-1.0 kg. The two cryostats contain 35 detectors with a total mass of 29.7 kg, fabricated with Ge material enriched to 88\% in isotope $^{76}$Ge, and 15.1 kg fabricated from natural Ge (7.8\% of $^{76 }$Ge). The 74.5\% yield of converting initial material into Ge diodes is the highest achieved to date. 
Starting from the innermost cavity, the cryostats 
are surrounded by an inner layer of electroformed copper, an outer layer of commercially obtained 
Oxygen-Free High thermal Conductivity (OFHC) copper, high-purity lead, an active muon 
veto, borated polyethylene, and polyethylene. The cryostats, copper, and lead shielding are all 
enclosed in a radon exclusion box that is purged with liquid nitrogen boil-off. The experiment 
is located in a clean room at the 4850-foot level (1478 m; 4300 m w.e.) of the Sanford Underground Research Facility (SURF) in Lead, South Dakota, USA.


The first results of the experiment (for statistics of 10 kg$\times$yr) were published in Ref. \cite{AAL17}. In these measurements, an unprecedented energy resolution was obtained for HPGe detectors (FWHM is 2.5 keV at $Q_{\beta\beta} = 2039 keV$) and a background level of 
$\sim 1.6\times10^{-3}$ c/keV$\times$kg$\times$yr was achieved. Limit on $0\nu\beta\beta$ decay of $^{76}$Ge was obtained as $T_{1/2} > 1.9\times10^{25}$ yr at 90\% C.L. (the Bayesian limit is 1.6$\times 10^{25}$ yr).
At the present level of background, the limit on $T_{1/2}$ is increasing nearly linearly and is projected to approach $\sim 1.5 \times 10^{26}$ yr for a 100 kg$\times$yr exposure.

\underline{The LEGEND experiment \cite{ABG17b}.} The LEGEND collaboration aims to increase the sensitivities for $^{76}$Ge in a first phase to $10^{27}$ yr and in a second phase up to $10^{28}$ yr. The goal is to perform a “background-free” measurement, defined as being $< 1$ mean expected background count at an experiments design exposure. 

LEGEND-200 plans to operate up to 200 kg of germanium detectors using the existing GERDA infrastructure at LNGS. In order to be “background-free” for an exposure of 1 t$\times$yr a factor of 5 reduction is needed relative to the latest GERDA and Majorana Demonstrator background levels. The existing infrastructure is large enough to house 200 kg of detectors: the neck of the cryostat has a diameter of 800 mm which is wide enough for 19 strings of detectors with a total outer diameter of 500-550 mm. LEGEND-200 will use the existing Majorana Demonstrator and GERDA P-PC and BEGe detectors as well as additional new detectors. The new ones will be of the inverted-coaxial type which offer a similar pulse shape performance but can have much higher mass. Thus the number of channels per kg and the resulting backgrounds from cables and holders will be reduced compared to the current experiments.

For the next phase of the experiment, LEGEND-1000, the exposure of 10 t$\times$yr is reached by operating 1000 kg of detectors for 10 years. This requires new infrastructure and a more ambitious background goal to remain in the background-free regime. Several options are still under consideration for LEGEND-1000, but an initial baseline design has been established with bare germanium detectors operating in liquid argon. Because the enrichment and detector production will be spread over several years, it is planned to install the detectors in several batches of $\sim$ 250 kg each. The data taking of the already installed detectors should continue largely undisturbed. 
The main cryostat volume is separated by thin copper walls from four smaller volumes of about 3 m$^3$ each. Each volume will house a subset of the detectors and is closed on top by a shutter. There will be a lock above each of the shutters such that the germanium detector array can be assembled in nitrogen atmosphere in a glove box together with the argon veto. An important design criterion will be the minimization of "dead" material, i.e. material like copper or teflon which contributes to the background and does not scintillate. One alternative design being considered is to use a scintillating plastic, such as polyethylene naphthalate (PEN) as a construction material since it has good mechanical properties. Compared to LEGEND-200 the background needs to be reduced by another factor of 6 (up to $\sim 3\times10^{-5}$ c/keV$\times$kg$\times$yr).


\subsubsection{Low temperature bolometer detectors}

The possibility of using low-temperature bolometers to search for a double beta decay was first proposed in 1982 by JINR (Dubna, Russia) researchers  G.V. Micelmacher, B.S. Neganov and V.N. Trofimov \cite{MIC82}. However, this idea was not developed at JINR. E. Fiorini learned about this proposal (see \cite{FIO15}) and already in 1984 E. Fiorini and T. Ninikovski published an article 
\cite{FIO84}, which detailed the prospects of using low-temperature detectors to search for double beta decay processes (with referring to work \cite{MIC82}). It was shown that energy resolution could be on the level of $\sim$ (0.1-0.2)\% (or even better!) at $\sim$ 2-3 MeV. And this technique was successfully developed by the efforts of the Milan group led by E. Fiorini. The TeO$_2$ crystal was used as the detector. The high content of $^{130}$Te in natural tellurium (34\%) made it possible to avoid the use of expensive enriched material and greatly simplified the work on growing crystals. Already in 1994, the first physical results  for a double beta decay of $^{130}$Te with a TeO$_2$ crystal (crystal weight was 330 g) were obtained  \cite{ALE94}. Then there were measurements with 4 crystals \cite{ALE96}, 20 crystals (MiBeta) \cite{ARN03} and 40 crystals (CUORICINO) \cite{AND11}. In the CUORICINO experiment, a sufficiently low value of the background index (BI = 0.18 c/keV$\times$kg$\times$yr) was achieved and strong limit on $0\nu\beta\beta$ decay of $^{130}$Te has been obtained, $T_{1/2} > 2.8 \times10^{24}$ yr. Then, the CUORE-0 experiment was implemented, in which the background level 0.06 c/keV$\times$kg$\times$yr was achieved \cite{ALD16}. 
All this led to the creation of a CUORE installation containing 988 TeO$_2$ crystals weighing 742 kg (see below). One of the main problems in this approach is a rather high level of background in the ROI. Basically, this background is associated with the decays of $\alpha$-particles located on the surface of crystals and surrounding structural materials. This problem was outlined at the measurement stage with the MiBeta, CUORICINO and CUORE-0 detectors. To solve this problem, the CUPID project \cite{WAN15,WAN15a} proposes the use of scintillating low-temperature bolometers. The technique of scintillating low-temperature bolometers has been successfully developing for the last 10-15 years. The idea is to simultaneously register both the thermal and light signals, which leads to a substantial suppression of the background from the $\alpha$-particles. Intensive investigations of the detecting properties of various crystals were carried out for the purpose of their application in experiments to search for neutrinoless double beta decay. Crystals grown from both natural and enriched materials ($^{100}$Mo, $^{82}$Se and $^{116}$Cd) were studied. The most promising at present are the crystals Li$_2$MoO$_4$, ZnSe, ZnMoO$_4$ and CdWO$_4$. In addition, the possibility of recording Cherenkov radiation in a TeO$_2$ crystal was studied and promising results have been obtained \cite{FAT10,BAT15}. All this allows us to hope for the successful implementation of projects such as CUPID and AMORE in the future. At present, the prototypes CUPID-0/Se \cite{ART17}, CUPID-0/Mo \cite{POD17} and AMORE-I \cite{JO17} are working (or ready to be started). Main goal of these prototypes is to demonstrate good enough parameters of using detectors and the possibility to reach background level of $\sim (10^{-3}-10^{-4})$ c/keV$\times$kg$\times$yr.
A lot of useful information about this technique one can find in the recent review \cite{POD17a}.

\underline{CUORE}. The CUORE (Cryogenic Underground Observatory for Rare Events) experiment \cite{ART15} is designed to search 
for $0\nu\beta\beta$  decay of $^{130}$Te using an array of TeO$_2$ cryogenic bolometers. The basic  working principle of CUORE is based on the calorimetric technique: the energy released in an absorber is measured through its temperature rise, which is read out by a sensitive thermal sensor attached to the absorber. 
CUORE is made of 988 natural TeO$_2$ cryogenic detectors, mounted in a cylindrical compact and granular structure of 19 towers, each made by 52 crystals arranged in 13 floors of 4 detectors each. The absorbers are cubic crystals ($5\times5\times5$ cm$^3$ each), with a total active mass of 742 kg (206 kg of $^{130}$Te). The temperature sensors that convert the thermal variation into an electrical signal are neutron transmutation doped (NTD) germanium semiconductor thermistors, glued on the absorber. The crystals are arranged in a copper structure, which serves as the heat bath.


 
The CUORE hut is built at the LNGS underground facility at an average depth of 3650 
m w.e.  A heavy shield consisting of layers of borated polyethylene, boric-acid powder, and lead bricks surrounds the cryostat to attenuate neutron and $\gamma$-ray backgrounds. More lead shielding is added inside the cryostat, including ancient Roman lead to further suppress the $\gamma$-rays from the cryostat materials. With a background index of 0.01 c/keV$\times$kg$\times$yr, the projected half-life sensitivity for CUORE is $9.5\times10^{25}$ yr (90\% C.L.), corresponding to an upper limit on the effective Majorana mass in the range of 50-200 meV, depending on the adopted NME calculation. 
CUORE is running experiment since April 2017. First results for 86.3 kg$\times$yr exposure have been presented in \cite{ALD17}. Achieved energy resolution is 7.7 keV (FWHM) and it is demonstrated that level of background in ROI region is $(0.014\pm 0.002)$ c/keV$\times$kg$\times$yr and new limit on $0\nu\beta\beta$ decay of $^{130}$Te is obtained as $ T_{1/2}> 1.3\times10^{25}$ yr (90\% C.L.). Combination with previous CUORICINO and CUORE-0 results gives $T_{1/2}> 1.5\times10^{25}$ yr (90\% C.L.). The median statistical sensitivity of this search is $\sim$ $0.8\times10^{25}$ yr (90\% C.L.) and more strong limit is obtained thanks to big "negative" fluctuation of the background. To be conservative I recommend to use exactly this last value as the most reliable and correct limit for $^{130}$Te by this moment.

\underline{CUPID} \cite{WAN15,WAN15a}. The CUPID (CUORE Upgrade with Particle IDentification)   
is a proposed future tonne-scale bolometric neutrinoless double beta decay experiment to probe the Majorana nature of neutrinos and discover Lepton Number Violation in the
so-called inverted hierarchy region of the neutrino mass. CUPID will be built on experience, expertise and lessons learned in CUORE, and will exploit the current CUORE infrastructure as much as
possible. In order to achieve its ambitious science goals, CUPID aims to increase the source mass and
dramatically reduce the backgrounds in the region of interest. This requires isotopic enrichment,
upgraded purification and crystallization procedures, new detector technologies, a stricter material selection, and possibly new shielding concepts with respect to the state of the art deployed in
CUORE. Main idea is to use scintillating bolometers based on scintillating crystals (as Li$_2$MoO$_4$, ZnSe, ZnMoO$_4$ and CdWO$_4$) or to use TeO$_2$ crystals with detecting of Cherenkov light. CUPID is developing in parallel of CUORE data taking and will start to work after finishing of the CUORE (probably in 2023-2025). 

\subsubsection{Loaded liquid scintillators}

The main idea of this approach is the use of existing large-scale low-background installations to search for $0\nu\beta\beta$ decay. Indeed, for the detection of solar and reactor neutrinos installations with extremely low background level at low energies were created (BOREXINO, KamLAND, SNO). Within the framework of these projects, phenomenal successes were achieved in the purification of various materials from radioactive impurities and the achievement of an extremely low background level. In 1994, R. Ragavan proposed to dissolve xenon in a liquid scintillator in order to search for 
$0\nu\beta\beta $ decay of $^{136}$Xe \cite{RAG94}. This proposal was made within the framework of the BOREXINO project, but it was not used. For the first time this idea was realized 15 years later in the KamLAND-Zen experiment (see below). In this case, an additional balloon with a liquid scintillator (approximately 13 tons) was placed in the center of the main KamLAND detector (1000 tons of liquid scintillator). And  the enriched $^{136}$Xe has been dissolved in the liquid scintillator of this balloon.
As a result, the liquid scintillator of the main detector served as a passive and active shield for the inner balloon. In 2005, within the framework of the SNO+ project, it was proposed to introduce into the liquid scintillator various substances containing nuclei of $\beta\beta$ decay isotopes. At first, the emphasis was on $^{150}$Nd, but after a while the Collaboration chose $^{130}$Te. In this case, it is proposed to introduce natural tellurium ($\sim$ 34\% of $^{130}$Te) into the liquid scintillator. Both these approaches (KamLAND-Zen and SNO+) are developing quite successfully now (KamLAND-Zen has been operating since 2011, and the start of data taking with SNO+ is expected in 2018). The positive side of this approach is the possibility of studying large amount of $\beta\beta$ isotopes and a very low background level, which is achieved "in a natural way", i.e., using existing low-background installations. The main drawback of this approach is not a very good energy resolution (currently it is $\sim$ 10\% FWHM at Q$_{\beta\beta}$). The shortcomings can also be attributed to a fairly narrow range of isotopes under study (up to now only $^{136}$Xe has been successfully studied). Although, in principle, it seems possible to study any isotopes in this approach .

\underline{KamLAND-Zen}. The detector KamLAND--Zen is a modification of the existing KamLAND 
detector carried out in the summer of 2011. The $\beta\beta$ source/detector is 13 tons 
of Xe-loaded liquid scintillator (Xe--LS) contained in a 3.08 m diameter 
spherical Inner Balloon (IB). The IB is suspended at the center of the KamLAND detector. 
The IB is surrounded by 1 kton of liquid 
scintillator (LS) contained in a 13 m diameter spherical Outer Balloon (OB) 
made of 135 $\mu$m thick composite film. The outer LS acts as an active shield 
for external $\gamma$-rays and as a detector for internal radiation from the Xe--LS or IB. 
The Xe--LS contains (2.44 $\pm$ 0.01) $\%$ by weight of enriched xenon gas (full 
weight of xenon is $\sim$ 320 kg). The isotopic abundances in the enriched xenon is
(90.93 $\pm$ 0.05) $\%$ of $^{136}$Xe. 
Scintillation light is detected by 1,325 17-inch and 554 20-inch photomultiplier tubes (PMTs). The energy resolution is 9.9 \% (FWHM) at 2.458 MeV. In phase-1 new limit on $0\nu\beta\beta$ decay of $^{136}$Xe has been obtained, $T_{1/2} > 1.9\times10^{25}$ yr (90\% C.L.) \cite{GAN13}.


KamLAND-Zen experiment phase-2 ~\cite{GAN16} (383 kg of the enriched xenon) has been finished in May 2016. The background index in the ROI on the level of $\sim 1.5\times 10^{-5}$ c/keV$\times$kg$\times$yr has been achieved ($\sim$ 30\% of the background is connected with 2$\nu$ events).
Limit on $0\nu\beta\beta$ decay of $^{136}$Xe of $9.2\times10^{25}$ yr (90\% C.L.) for 534.5 days of measurements has been established. Combining this limit with result of a phase-1  authors obtained limit of $1.07\times10^{26}$ yr (90\% C.L.), which corresponds to limit $\langle m_{\nu} \rangle$ $< 0.06-0.16$ eV. It must be emphasized that sensitivity of the experiment is $5\times10^{25}$ yr ($\langle m_{\nu} \rangle$ $\sim 0.09-0.24$ eV), and more strong limit is obtained thanks to big "negative" fluctuation of the background. To be conservative I recommend to use exactly this last value as the most reliable and correct limit by this moment. Nevertheless for today this is the best limit on $\langle m_{\nu} \rangle$ among all experiments. 
     Now there is a preparation of measurements with 750 kg of the enriched xenon and a new (pure) internal balloon and sensitivity of the experiment will be increased up to $\sim 2\cdot10^{26}$ yr.
     
In the future it is planned to reduce $2\nu\beta\beta$ background by improving the energy resolution
by a factor of 2 to get $\Delta$E/E (FWHM) $\sim$ 4.7\% at the Q$_{\beta\beta}$-value ($\sim$ 2.5 MeV). Various R\&D works are underway for better energy resolution (LAB-based LS with higher light yield, high quantum-efficiency PMTs with light collectors), background rejection by new LS purification method, an imaging sensor system for
 $\beta$-$\gamma$ discrimination and a scintillating balloon for removal of $^{214}$Bi background. A pressurized Xe-LS environment for increased amount of Xe up to 1000 kg is also considered. The plan called KamLAND2-Zen aims to make the $\langle m_{\nu}\rangle$ sensitivity to cover the region of inverted hierarchy \cite{SHI17}.      

\underline{SNO+}~\cite{AND16,ARU17}. SNO+ is a multipurpose liquid scintillator neutrino experiment based at SNOLAB
in Sudbury, Ontario, Canada. The experiment's main physics goal is a search for neutrinoless
double beta decay in $^{130}$Te. SNO+ reuses the existing infrastructure of SNO, consisting of a 6 m radius acrylic vessel surrounded by almost 10,000 photomultiplier tubes. A depth of 6000 m w.e. provides shielding from cosmic muons. In SNO+, the acrylic vessel will be filled with liquid
scintillator (780 tons) and surrounded with a shielding of ultra-pure water. In Phase I of the experiment high sensitivity to neutrinoless double beta decay will be pursued by mixing 3.9 tonnes (0.5\% by weight) of natural Te into the scintillator. The energy resolution will be 10.8 \% (FWHM) at 2.5 MeV. Primary goal is the search for neutrinoless double beta decay of $^{130}$Te with expected sensitivity of $1.9\times 10^{26}$ y (90\% C.L.). The possibility of deploying up to ten times more of natural tellurium has been investigated, which would enable SNO+ to achieve much better sensitivity in the future.

Significant work has taken place to transform the heavy water detector of SNO into a liquid scintillator detector.  The neutrinoless double beta decay phase will begin in late 2018. 

\subsubsection{Liquid and gas Xe TPC}

The first $\beta\beta$ decay measurements with $^{136}$Xe were performed using a scintillation detector based on liquid Xe \cite{BAR86} and a high-pressure gas ionization chamber \cite{BAR89}. In the early 90's, measurements using a proportional counter \cite{BEL91} and a gas TPC~\cite{WON91,ART92} were realised. In 2000, the EXO experiment \cite{DAN00} was presented in which it was proposed to register all the products of the $\beta\beta$-decay of $^{136}$Xe (including registration of the $^+$Ba ion, the daughter nucleus after the $\beta\beta$ decay of $^{136}$Xe) using gas (or liquid) TPC\footnote{Original idea of such approach first time was proposed by M. Moe in 1991 \cite{MOE91}.}. It was quite difficult to organize the registration of the Ba ion under real experiment conditions, so at the initial stage it was decided to abandon this option. As a result, a prototype EXO-200 was created. In this version, a liquid-xenon TPC is used and only electrons from $\beta\beta$ decay are recorded. Moreover, both ionization and scintillation signals are simultaneously recorded, which allows achieving a reasonably good energy resolution ($\sim$ 2.9\% FWHM at 2.5 MeV). EXO-200 is an operational experiment in which good physical results are obtained and which is a prototype of the next-generation nEXO experiment (see below for more details). Another promising direction is a high-pressure gas TPC (NEXT \cite{FER17} and PandaX-III \cite{CHE17} experiments). In this case, the emphasis is on a very good energy resolution ($\sim$ 1\%) and the possibility of restoring electron tracks from double beta decay. Both experiments are at the stage of creation and research of prototypes. Here it should be noted that the gas option is also considered for the EXO approach.

\underline{EXO-200}. The experiment is located at the Waste Isolation Pilot
Plant (WIPP) in Carlsbad, NM, USA (1624 m w.e.). The EXO-200 detector is described in detail 
in Ref. \cite{AUG12}. 
The central component of the detector is a 
LXe time projection chamber (TPC). The chamber is divided into two equal volumes by a photoetched phosphor bronze cathode plane. Near each end of the chamber are copper platters for housing an array of large-area avalanche photodiodes (LAAPDs) and two wire planes crossed at 60$^{\circ}$. The wire signals provide two-dimensional position information of the event. The position in the drift direction can be obtained using the known drift speed and the light signal collected by the APDs as time zero. Full mass of enriched Xe is 175 kg and mass in the active volume is ~$\sim$ 110 kg.
EXO-200 began low-background data taking in June 2011. The detector successfully operated for two and half years before underground fire and radiological accidents at WIPP. Using the first two years of data, the collaboration published a $0\nu\beta\beta$ decay search result with total $^{136}$Xe exposure of 100 kg$\times$yr \cite{ALB14}. Level of background at ROI was 
$1.7\times10^{-3}$ c/keV$\times$kg$\times$yr. Limit on $0\nu\beta\beta$  decay of $^{136}$Xe was established as 
$T_{1/2}(0\nu\beta\beta) > 1.1\times10^{25}$ yr at 90\% C.L. In addition to the $0\nu\beta\beta$ decay search, EXO-200 has published a precision measurement of $2\nu\beta\beta$  decay of $^{136}$Xe, $T_{1/2}(0\nu\beta\beta) = 2.165 \pm 0.016 (stat) \pm 0.059 (sys)\times 10^{21}$ yr\cite{ALB14a}. This has been the most precisely measured $2\nu\beta\beta$ decay half-life of any isotope.
EXO-200 new low background data taking began on April 29, 2016 and is expected to continue until 2018 (Phase-II). Sensitivity studies indicate that with the detector upgrades and analysis improvements, EXO-200 can reach a $0\nu\beta\beta$ decay half-life sensitivity of $5.7\times10^{25}$ yr, using combined Phase-I and Phase-II data. 


\underline{The nEXO experiment \cite{ALB17}}. 
The success of EXO-200 demonstrates that LXe TPC technology is well suited for a large-scale 
double beta decay experiment. nEXO is a proposed $\sim$ 5-tonne detector. Its design will be optimized 
to take full advantage of the LXe TPC concept and can reach $0\nu\beta\beta$ half-life sensitivity of $\sim 6\times 10^{27}$ yr (for five years of measurements). 
A substantial R$\&$D program is underway to validate the nEXO detector design concepts and 
technologies. After a 1-2 year of R$\&$D phase, nEXO will be ready to move towards detector conceptual design. As an upgrade option for nEXO, technologies that can tag the $\beta\beta$ daughter barium nuclei efficiently in situ are under development. Such a Ba-tagging technique, if realized, can help with decreasing of a background and with increasing of a sensitivity. Since Ba-tagging is still in early R$\&$D, it will not be incorporated in the initial phase of nEXO.
The experiment looks very promising. Main problem here is possibility to produce 5 tons of enriched Xe (see remark in section 5.4).

\subsubsection{Tracking detectors}

The main idea of ​​this approach is the visualization of the tracks of electrons and obtaining complete information about the $\beta\beta$ event: the total decay energy, the energy of each electron, the angle between electrons, the simultaneity of the emission of the electrons, and the observation of the event vertex. All this information allows not only to distinguish useful events from background ones, but also to study the parameters of $\beta\beta$ processes. In particular, this made it possible to confirm 1$^+$ state dominance hypothesis in the $2\nu\beta\beta$ decay of $^{100}$Mo by investigating the shape of the spectrum of single electrons in the NEMO-3 experiment. In the case of $0\nu\beta\beta$ decay, this will allow, for example, to distinguish the mass decay mechanism from the case of "right-hand currents" (by examining the spectrum of single electrons and the angular distribution of electrons). The track approach in double beta decay began with the use of the Wilson cloud-chamber in 1951~\cite{LAW51}. Then C. Wu and co-authors used a "track" detector to search for the neutrinoless double beta decay of $^{48}$Ca \cite{BAR70} and $^{82}$Se \cite{KLE75}. A thin planar target containing the investigated isotope was placed in a streamer chamber and surrounded on both sides by scintillation counters. Plastic scintillators provided information on the simultaneity of the emission of the electrons and their energy. Photos from the streamer camera gave information about the tracks of electrons and their energy (by the curvature of the tracks). Almost the same scheme was used in NEMO detectors (NEMO-2~\cite{ARN95} and NEMO-3~\cite{ARN05}). But instead of a streamer camera, Geiger counters were used, which made it possible to switch to electronic recording of events. One of the advantages of this technique is the possibility of studying almost all possible $\beta\beta$ candidates (with the exception of $^{136}$Xe). So in the NEMO-3 experiment, seven different isotopes were simultaneously investigated! The disadvantages of the method include not very good energy resolution ($\sim$ 4-8\% FWHM at 3 MeV) and rather low registration efficiency of events ($\sim$ 10-30\%).

It should be noted that information on electron tracks can also be obtained in compressed gas xenon TPC (NEXT, PandaX-III, gaz EXO).

\underline{NEMO-3}.
The objective of the NEMO-3 experiment was the search for the $0\nu\beta\beta$  decay and investigation of the $2\nu\beta\beta$  decay. Its method is based on the detection of electrons in the tracking device and the energy measurement in calorimeter. The NEMO-3 detector has a cylindrical shape and is composed of twenty equal sectors. It contains almost 9 kg of seven different $\beta\beta$ isotopes in the form of thin ($\sim$ 50 mg/cm$^2$ ) source foils located vertically in the middle of tracking volume surrounded by a calorimeter. The tracking chamber is made of 6180 open octagonal drift cells operating in Geiger mode. The position resolution for tracking is 0.6 mm in the horizontal plane and 0.8 cm in the vertical direction. The calorimeter comprises 1940 plastic scintillator blocks coupled to low radioactivity photomultipliers (PMT). For 1 MeV electrons the timing resolution is 250 ps and the energy resolution (FWHM) is about 8\% at 3 MeV. A 25 Gauss magnetic field is used for electron-positron discrimination by the track curvature. The detector is shielded from external gamma rays by 18 cm of low activity iron and against neutrons by 30 cm of borated water. 

The detector is capable of identifying $e^-$ , $e^+$ , $\gamma$ and $\alpha$ particles and allows a good discrimination between signal and background events. The full event kinematics reconstruction available with the NEMO-3 track-calorimetric approach is useful for the study of the underlying $\beta\beta$ decay mechanism. 

The detector was operating in the Modane underground laboratory located in the Frejus tunnel at depth of 4800 m w.e. The data taking started in February 2003 is completed in January 2011. In total 
$1.15\times10^9$ events have been triggered in 6391 runs dedicated to $\beta\beta$ decay study during 6.1 yr of effective data taking. 
Events are selected requiring two reconstructed electron tracks originating from the common vertex in the source foil. The cut on the minimal energy deposited by each electron in calorimeter of 200 keV is used. The detection efficiency of the $0\nu\beta\beta$ decay events is $\sim$ 10\%. A full description of the detector and its characteristics can be found in~\cite{ARN05}.

Using the NEMO-3 detector, seven isotopes have been investigated simultaneously ($^{48}$Ca, $^{82}$Se, $^{96}$Zr, $^{100}$Mo, $^{116}$Cd, $^{130}$Te and $^{150}$Nd). For all seven isotopes, a search for two-neutrino decay, neutrinoless decay and decays with the Majoron emission has been performed. As a result, half-life values of $2\nu\beta\beta$ decay have been measured for all seven isotopes \cite{ARN10,BAR11,ARN11,ARN15,ARN16,ARN16b,ARN17}. In addition $2\nu\beta\beta$ decay of $^{100}$Mo to the 0$^+_1$ excited state of $^{100}$Ru has been detected too \cite{ARN07}. In each case, the main decay parameters were measured: the total energy spectrum, the spectrum of single electrons, and the angular distribution of electrons. 

Limits on neutrinoless double beta decay and on decays with the Majoron emission for all seven isotopes  have also been established \cite {ARN10,BAR11,ARN11,ARN15,ARN16,ARN16b,ARN06,ARN17}. In particular, for $^{100}$Mo, the limit $T_{1/2}(0\nu)> 1.1\times10^{24}$ yr ($\langle m_{\nu} \rangle$  $< 0.33-0.62$ eV)  has been achieved \cite{ARN15}. The level of observed background in the $0\nu\beta\beta$ signal region [2.8-3.2] MeV is $0.44\pm 0.13$ c/kg$\times$yr ($\sim$ 50\% of the background is connected with 2$\nu$ events) and no events are observed in the interval [3.2-10] MeV.


\underline{SuperNEMO}.
SuperNEMO aims to extend and improve the successful NEMO-3 technology. It will extrapolate 
NEMO-3 by one order of magnitude by studying about 100-140 kg of $\beta\beta$ isotope(s). The 
detector’s ability to measure different $\beta\beta$ isotopes and reconstruction of the topological signature of the decay are distinct features of SuperNEMO. The baseline isotope choice for SuperNEMO is $^{82}$Se. However other isotopes are possible. In particular, $^{150}$Nd and 
$^{48}$Ca are being looked at. Detector will be able to measure individual electron tracks, vertices, energies and time of-flight, and to reconstruct fully the kinematics and topology of an event. Particle identification of gamma and alpha particles, as well as distinguishing electrons from positrons with the help of a magnetic field, form the basis of background rejection. An important feature of NEMO-3 which is kept in SuperNEMO is the fact that the double beta decay source is separate from the detector, allowing several different isotopes to be studied. SuperNEMO will consist of about twenty identical modules, each housing around 5-7 kg of isotope. The project is completing a 3 year design study and R$\&$D phase with much progress towards the first prototype Demonstrator module. The R$\&$D program focuses on four main areas of study: isotope enrichment, tracking detector, calorimeter, and ultra-low background materials production and measurements. 
The source is a thin ($\sim$ 40-50 mg/cm) foil inside the detector. It is surrounded by a gas tracking chamber followed by calorimeter walls. The tracking volume contains more than 2000 wire drift chambers operated in Geiger mode, which are arranged in nine layers parallel to the foil. The calorimeter is divided into 712 blocks which cover most of the detector outer area and are read out by photomultiplier tubes (PMT). Planed energy resolution is 4\% (FWHM) at 3 MeV and planed detection efficiency for $0\nu\beta\beta$ decay is $\sim$ 30\%. More detailed information one can find in \cite{BAR17,CAS16}.
The SuperNEMO Demonstrator should start data taking in 2018.


\subsection{Background}

One of the main problem in experiments on double beta decay is the background problem. It is the background events in the ROI that limit the sensitivity of the experiments. To reduce the background associated with cosmic rays, the experimental setups, as a rule, are located deep underground. To suppress the background associated with natural radioactivity, the detectors are surrounded by layers of passive and active shields. The detectors themselves are made of materials with an extremely low content of radioactive impurities. The main problem is the $^{214}$Bi (Q$_{\beta}$ = 3.27 MeV) and $^{208}$Tl (Q$_{\beta}$ = 4.99 MeV) isotopes, which are the daughter products of the decay in the  $^{238}$U and $^{232}$Th chains, respectively. Typically, clean construction materials are checked for radioactive impurities using low-background HPGe detectors. This method allows to measure the activity of $^{214}$Bi and $^{208}$Tl at a level up to $\sim$ 10-100 $\mu$Bq/kg (see, for example, \cite{BUD08}). A BiPo method is based on measuring the $\beta-\alpha$ coincidence of $^{214}$Bi-$^{214}$Po and $^{212}$Bi-$^{212}$Po. The created installation BiPo-3 \cite{BAR17a} allows to measure the activity of $^{214}$Bi and $^{208}$Tl in thin samples at a level of $\sim$ 10 and 2 $\mu$Bq/kg, respectively. The use of mass-spectrometric methods allows to determine with high sensitivity ($\sim 10^{-12}- 10^{-14}$ g/g~\cite{ABG16}) the concentration of $^{238}$U and $^{232}$Th only. If the decay products are in equilibrium, this gives information about the activity of $^{214}$Bi and $^{208}$Tl. Unfortunately, very often the equilibrium is broken and, then, it is necessary to bear this in mind when determining the activity of $^{214}$Bi and $^{208}$Tl.

Thus, in order to create low-background conditions in experiments to search for neutrinoless double beta decay, it is necessary:

1) to place the detector deep underground;

2) to use passive and active shields;

3) to manufacture passive shield and design elements of the detector from low-background materials;

4) to use the investigated $\beta\beta$ isotopes with an extremely low content of radioactive impurities.

In addition, it is very important to suppress the background using the characteristics of the detectors themselves. The most important parameter here is the energy resolution of the detector - a good energy resolution allows significantly narrowing the ROI and, thereby, reducing the number of background events in it. Among currently used detectors, HPGe detectors and low-temperature bolometers  have the best parameters (FWHM is $\sim$ 0.1\% and $\sim$ 0.2\% respectively at energy $\sim$ 2-3 MeV). One of the worst (in terms of resolution) is detectors based on liquid scintillators (FWHM is $\sim$ 10\% at energy $\sim$2-3 MeV). Important parameters for background suppression are also the granularity of the detector, the possibility of discrimination of background events using information about the shape of a pulse, the ability to discriminate internal Bi-Po events, the ability to record electron tracks and reconstruct the vertex of $\beta\beta$ events.
Table 6  shows the values of the background index for the best experiments performed so far. 

\begin{table}[ht]
\caption{The background index (BI) and sum background in ROI ($\Sigma$B) of the current $\beta\beta$ decay experiments. M is mass of the investigated isotope; Q$_{\beta\beta}$ is the energy of $\beta\beta$ decay; $\Delta$E/E is the energy resolution (FWHM) at Q$_{\beta\beta}$. $\Sigma$B is given for 5 years of measurement. For KamLAND-Zen results for 1 m radius detector are presented. $^{a)}$ Total weight of the registration part of the detector (which is used to calculate the BI).}
\hspace{0.2cm}
\begin{center}
\begin{tabular}{llllll} 
\hline
Experiment & Isotope & M,  & $\Delta$E/E,  & BI,  &
$\Sigma$B  \\ 
& & kg & \% at Q$_{\beta\beta}$ & c/kev$\times$kg$\times$yr & (ROI = $\Delta$E) \\ 
\hline
CUORE~\cite{ALD17} & $^{130}$Te  & 206 (742)$^{a)}$ & 0.32 & 0.014  & 400 \\
GERDA-II~\cite{AGO17a} & $^{76}$Ge & 35 & 0.14 & 0.001 & 0.5\\
Majorana-D~\cite{AAL17} & $^{76}$Ge & 30 & 0.12 & 0.0016 & 0.6 \\
EXO-200~\cite{ALB14} & $^{136}$Xe & 100 & 2.9 & 0.0015 & 55 \\
KamLAND-Zen~\cite{GAN16} & $^{136}$Xe & 112 ($3.9\times 10^3$)$^{a)}$ & 10 & $\sim 1.5\times 10^{-5}$ & $\sim$ 70 \\
NEMO-3~\cite{ARN15} & $^{100}$Mo  & 7 & 8 & 0.0011 & 9 \\  
\hline
\end{tabular} 
\end{center}
\end{table}

\section{Main features of isotopes to investigate double beta decay}

The use of enriched isotopes in experiments on the search for double beta decay began with the first measurements. Already in the first experiment in 1948 \cite{FIR48,FIR49}, a sample of $^{124}$Sn with a weight of 25 g and an enrichment of 50\% was studied. For a long time, a few grams (tens of grams) of the isotope were sufficient for the experiments. In the famous M. Moe experiment \cite{ELL87}, in which for the first time in a direct experiment a two-neutrino double beta decay was recorded, only 14 g of the enriched $^{82}$Se were used. Therefore, the question of the isotopes price was not very important. Beginning in the late 1980s, enriched isotopes began to be studied in an amount of 
$\sim$ 0.1-10 kg, at present experiments are under way using $\sim$ 30-300 kg of isotopes, and for the next generation of experiments it is planned to use $\sim$ 1000- 5000 kg of enriched isotopes. In order to check the area of ​​the normal neutrino hierarchy, tens of tons of enriched isotopes will be required (see section 5). Therefore, there are important questions about the possibility of obtaining the necessary amount of isotopes and their price. We emphasize, however, that in order to investigate the two-neutrino decay it is sufficient to use, as a rule, $\sim$ 0.1-1 kg of isotope (in this case the price is not so important). The enriched isotopes that have been used so far have been produced mainly in Russia (or in the Soviet Union) and in small quantities in the United States. Recently, for the LUCIFER experiment \cite{ART17}, 
$\sim$ 15 kg of enriched $^{82}$Se were produced in the European Union (URENCO, Netherlands). The usual level of enrichment is $\sim$ 90-95\%, since the price of isotopes with an enrichment level of 99\% or more sharply increases.

\subsection{Possibilities for double beta isotope production}

There are different methods for isotope production as follows:  

- centrifugal separation (productivity in arbitrary units is 1); 

- laser separation (productivity is $\sim$ 0.1); 

- plasma separation (productivity is $\sim$ 0.01); 

- electromagnetic separation (productivity is $\sim$ 0.001). 

Taking into account the productivity and cost (which is inversely proportional to the productivity), it is 
clear that the centrifugal separation is the only method which could produce 0.1-10 tons of enriched 
material for the $\beta\beta$-decay experiments. Present total productivity can be estimated as $\sim$ 200 kg 
per year. It can be increased by $\sim$ 10 times with additional financial investment. If necessary, a new facility could be organized for this goal too. 
There are certain requirements concerning the feasibility of compounds for centrifugal separation of stable isotopes. Compounds that are gaseous at ambient temperature are most suitable. Secondly, the sublimation temperature of a compound in vacuum must be lower than its decomposition temperature. The compound should not be corrosive or react with the material of the rotor of the gas centrifuge. Note that at present among the above-mentioned isotopes only $^{76}$Ge, $^{82}$Se, $^{100}$Mo, $^{116}$Cd, $^{124}$Sn, $^{130}$Te and $^{136}$Xe can be produced (and have been produced) by centrifugation. There is a hope to produce $^{150}$Nd by centrifugation too (if some technical problems will be solved). The method still cannot be applied to $^{48}$Ca and $^{96}$Zr. The $^{48}$Ca, $^{96}$Zr and $^{150}$Nd samples (with mass $\sim$ 10-50 g and enrichment $\sim$ 50-90\%) used up to now in various experiments were produced by the electromagnetic separation method. In Table 7  one can see total amount of isotopes, produced for double beta decay experiments during a last few decades. Note that very recently $\sim$ 110 kg of $^{100}$Mo were produced in Russia for the AMORE experiment and $\sim$ 150 kg of $^{76}$Ge will be produced in Russia during nearest few years for the LEGEND experiment. So, really we are entering a new era of experiments with a few hundred kg of isotopes.

\begin{table}[ht]
\caption{Approximate amount of $\beta\beta$ decay isotopes have been produced up to now.}
\begin{center}
\begin{tabular}{lllllllllll} 
\hline
Isotope & $^{48}$Ca & $^{76}$Ge  & $^{82}$Se  & $^{96}$Zr & $^{100}$Mo & $^{116}$Cd & $^{124}$Sn & $^{130}$Te & $^{136}$Xe & $^{150}$Nd  \\ 

M, kg & 0.03  & 120 & 22 & 0.03 & 125 & 1.5 & 0.5 & 25 & 1000 & 0.07 \\  
\hline
\end{tabular} 
\end{center}
\end{table}

\subsection{Price of isotopes}

\begin{table}[ht]
\caption{Approximate price of $\beta\beta$ isotopes obtained by 
centrifugation. $^{a)}$Taking into account possible 20\% price reduction in the mass production case.}
\hspace{0.2cm}
\begin{center}
\begin{tabular}{llll} 
\hline
Isotope & Abundance & Price per kg (k\$) & Cost of 10 tons (Mln.\$) \\  
\hline
$^{76}$Ge & 7.61  & $\sim$ 80 & 800 (640)$^a)$ \\
$^{82}$Se & 8.73  & $\sim$ 80 & 800 (640)$^a)$ \\
$^{100}$Mo & 9.63 & $\sim$ 80 & 800 (640)$^a)$ \\
$^{116}$Cd & 7.49  & $\sim$ 180  & 1800 (1440)$^a)$ \\
$^{124}$Sn & 5.79  & $\sim$ 300  & 3000 (2600)$^a)$  \\
$^{130}$Te & 34.08  & $\sim$ 20 & 200 (160)$^a)$ \\
$^{136}$Xe & 8.87 & $\sim$ 5-10 & 50-100 (40-80)$^a)$ \\
$^{150}$Nd & 5.6  & $> 300$ & $> 3000$ \\ 
\hline
\end{tabular}
\end{center}
\end{table}

As already noted, when using ever larger and larger isotope masses in modern experiments, one of the limiting factors is the price of the isotope. Table 8 shows the approximate prices of isotope candidates for double beta decay produced by centrifugation. 
When compiling the table, information on the production of enriched isotopes for previous experiments was used. It is clear that it is necessary to treat these values carefully (just as preliminary estimation), and, in addition, prices may change with time. It is seen from the table that  $^{136}$Xe and  
$^{130}$Te are the cheapest isotopes. In the case of $^{136}$Xe, this is due to the fact that we are dealing with a gas and the stages of transferring matter to the gaseous state and then back do not exist (the so-called "chemistry" steps are not required). In the case of $^{130}$Te, this is due to the high content of the $^{130}$Te in the starting material (natural tellurium contains 34\% of $^{130}$Te). $^{150}$Nd has not been obtained to date by the centrifuge method, but there are certain developments that allow us to hope for its production by centrifugation in the future. This requires special equipment, etc. Therefore, it is clear that if this opportunity will be realized, the price for $^{150}$Nd will be quite high (in Table 8 just preliminary estimation is presented).

\subsection{Specificity of using isotopes in various experiments}

As a result of separation by the centrifugation method, the content of the most dangerous radioactive impurities ($^{214}$Bi and $^{208}$Tl) in the samples obtained usually is at the level of a few mBq/kg. Apparently, this is determined by the purity of chemical reagents used in the final stage of transfer of the isotope from the gaseous state to the solid state\footnote{Monitoring the purity of the reagents used can help reduce the concentration of impurities. In the production of $^{82}$Se for the LUCIFER experiment, careful monitoring of the reagents used was carried out, which allowed obtaining enriched selenium with a low content of radioactive impurities ($^{214}$Bi $<0.11$ mBq/kg and $^{208}$Tl $<0.04$ mBq/kg \cite{ART17})}. Therefore very often produced isotopes are subjected to additional purification before use in experiments. Methods of chemical and "physical" purification are used. In the latter case, it is a vacuum distillation or purification during the crystallization process. For example, in the NEMO-3 experiment, a portion of the enriched $^{100}$Mo was purified by the chemical method \cite{ARN05,ARN01}, and part by the physical method (by growing a monocrystal of $^{100}$Mo in a vacuum) \cite{ARN05}. After purification by the chemical method, the activity of $^{214}$Bi and $^{208}$Tl was $\sim$ 0.3 and $\sim$ 0.1 mBq/kg, and after the physical purification $\sim$ 0.06 and $\sim$ 0.09 mBq/kg, respectively \cite{ARN15}. In the NEMO-3 experiment, part of $^{116}$Cd was purified by distillation in vacuum, and for the SuperNEMO-Demonstrator experiment, $^{82}$Se was purified using double distillation methods ($^{208}$Tl activity after purification was $\sim$ 0.02 mBq/kg \cite{BAR17a}). When the enriched isotope is used as the working substance of crystalline detectors (HPGe, TeO$_2$, CdWO$_4$, Li$_2$MoO$_4$, ZnMoO$_4$, ...), impurities (including the radioactive ones) are purified by preparing the components of the used charge, and then, "automatically", in the process growth of the corresponding crystals. As a result, it is possible to achieve the activity of $^{214}$Bi and $^{208}$Tl in the crystals at a level of a few $\mu$Bq/kg \cite{ARM17}. It was also demonstrated (by the example of CdWO$_4$) that re-crystallization allows to reduce the content of radioactive impurities in the crystals \cite{BAR16}. In the case of $^{136}$Xe, "chemistry" is not required at the final stage of the production, so there is no pollution at this stage. And if the cylinders for storing xenon are properly prepared, then problems with the content of radioactive impurities should not exist. In addition, xenon can be easily purified from impurities using various getters and by distillation. In the case of HPGe crystals, the initial germanium passes through several stages of deep purification from impurities (including zone melting), then a "natural" purification from impurities during crystal growth occurs. As a result, HPGe detectors "automatically" have a very low content of impurities (including radioactive impurities). Thus, it was shown that the content of $^{238}$U and $^{232}$Th in HPGe crystals is less than $2\times10^{-15}$ g/g \cite{MAJ03}. In a few times stronger limit was obtained recently using GERDA-I data \cite{AGO17b}.

\section{Experimental approaches in the case of a normal hierarchy}

If we are dealing with a normal hierarchy, we will need to investigate the region of $\langle m_{\nu} \rangle$ $\approx$ (1-5) meV\footnote{I discuss here the limiting case when mass of lightest neutrino $m_0$ is lower than 1 meV. If $m_0$ is higher then $\langle m_{\nu} \rangle$ could be higher too (up to $\sim$ 30 meV) - see Fig. 2 from Ref. \cite{SIM16}. Follow \cite{SIM16} allowed region for $m_0$ is $<$ 26 meV in case of normal hierarchy and $<$ 8 meV in case of inverted hierarchy.}. For the sake of simplicity, let us consider the possibility of realizing an experiment with 10 tons of an isotope. Table 9 presents the number of nuclei contained in 10 tons of different isotopes. 

\begin{table}[ht]
\caption{The number of nuclei in 10 tons of isotope and the number of decays after 10 years of
measurement (for $T_{1/2} = 10^{29}$ yr). Q$_{\beta\beta}$ is the energy of the $\beta\beta$ decay.}
\hspace{0.2cm}
\begin{center}
\begin{tabular}{llll}
\hline
Isotope & Q$_{\beta\beta}$, & No. of nuclei & No. of decays in 10 tons \\ 
 & keV & in 10 tons of isotope & during 10 yr ($T_{1/2} = 10^{29}$ yr) \\  
\hline
$^{48}$Ca & 4267.98  & $1.25\times 10^{29}$ & 8.6 \\
$^{76}$Ge & 2039.00 & $7.9\times 10^{28}$  & 5.5   \\
$^{82}$Se & 2997.9  & $7.3\times 10^{28}$  & 5 \\
$^{96}$Zr & 3355.85  & $6.3\times 10^{28}$  & 4.3 \\
$^{100}$Mo & 3034.40 & $6\times 10^{28}$ & 4.1 \\
$^{116}$Cd & 2813.50  & $5.2\times 10^{28}$ & 3.6 \\
$^{124}$Sn & 2292.64& $4.8\times 10^{28}$ & 3.3 \\
$^{130}$Te & 2527.52  & $4.6\times 10^{28}$ & 3.2 \\
$^{136}$Xe & 2457.83 & $4.4\times 10^{28}$ & 3 \\
$^{150}$Nd & 3371.38  & $4\times 10^{28}$ & 2.8 \\ 
\hline
\end{tabular}
\end{center}
\end{table}

\begin{table}[ht]
\caption{Half-life values (in yr) for different values of $\langle m_{\nu} \rangle$. The same NMEs and $g_A$ value as in Table 2 were used. The bold type denotes values achievable in measurements with a 10 ton detector (10 yrs of measurement).}
\hspace{0.2cm}
\begin{center}
\begin{tabular}{llll} 
\hline
Isotope & $\langle m_{\nu} \rangle$ = 1 meV & $\langle m_{\nu} \rangle$ = 3 meV & 
$\langle m_{\nu} \rangle$ = 5 meV \\   
\hline
$^{48}$Ca & $(0.58-14)\times 10^{30}$  & $({\bf 0.64}-16)\times 10^{29}$ & $({\bf 0.23}-5.6)\times 10^{29}$ \\
$^{76}$Ge & $(1.2-7.7)\times 10^{30}$  & $(1.3-8.5)\times 10^{29}$ & $({\bf 0.48}-3.1)\times 10^{29}$ \\
$^{82}$Se & $(3.6-20.1)\times 10^{29}$  & $({\bf 0.4}-2.3)\times 10^{29}$  & $({\bf 1.4-8.3})\times 10^{28}$ \\
$^{96}$Zr &  $(1.3-11)\times 10^{29}$ & $({\bf 1.4}-12)\times 10^{28}$  & $({\bf 0.5-4.4})\times 10^{28}$ \\
$^{100}$Mo & $(1.2-4.2)\times 10^{29}$ & $({\bf 1.3-4.7})\times 10^{28}$ & $({\bf 0.5-1.7})\times 10^{28}$ \\
$^{116}$Cd & $(1.9-6.1)\times 10^{29}$ & $({\bf 2.1-6.8})\times 10^{28}$ & $({\bf 0.8-2.5})\times 10^{28}$ \\
$^{124}$Sn & $(0.4-2.5)\times 10^{30}$ & $({\bf 0.45}-2.8)\times 10^{29}$ & $({\bf 1.6-10})\times 10^{28}$ \\
$^{130}$Te & $(0.28-4.0)\times 10^{30}$  & $({\bf 0.3}-4.4)\times 10^{29}$ & $({\bf 0.1}-1.6)\times 10^{29}$ \\
$^{136}$Xe &$(0.4-2.9)\times 10^{30}$ & $({\bf 0.4}-3.2)\times 10^{29}$ & $({\bf 0.16}-1.2)\times 10^{29}$ \\
$^{150}$Nd & $({\bf 0.5}-5.7)\times 10^{29}$ & $({\bf 0.6-3.3})\times 10^{28}$ & $({\bf 0.2-2.3})\times 10^{28}$ \\ 
\hline
\end{tabular}
\end{center}
\end{table}

In addition, the number 
of events obtained with 10 tons of isotope after 10 years of measurement with 100\% efficiency 
and for $T_{1/2} = 10^{29}$ yr are shown. It is possible to see that a small number of useful events ($\sim$ 3-9 events) in this case are expected. For more realistic efficiency one can wait for not more then 2-6 events. So, one can conclude that effect can be registered only if $T_{1/2} \le 10^{29}$ yr and background is close to zero. In Table 10 the estimated half-life values for different isotopes and for $\langle m_{\nu}\rangle$ = 1, 3 and 5 meV are presented. These estimations were done using recent (most reliable) NME calculations and in assumption that $g_A$ = 1.27. $T_{1/2}$ values at which the $0\nu\beta\beta$ decay could be registered (in principle), are shown in bold.
Using the information from Table 9  and Table 10  one can conclude that with 10 tons of isotope, the sensitivity to $\langle m_{\nu}\rangle$ at the level 3-5 meV can be reached with many isotopes, but it is practically impossible to reach 1 meV sensitivity. To completely cover 1 meV region one will need as minimum 30-100 tons of isotope. The best sensitivity can be reached with $^{100}$Mo, $^{116}$Cd and $^{150}$Nd. In case of $^{76}$Ge it will be difficult to reach even 5 meV sensitivity using 10 tons detector.

\subsection{Possible experimental approaches}

Let us consider some possible experimental approaches to such measurements: 

- HPGe detectors; 

- low temperature bolometers; 

- liquid scintillation detectors (using KamLAND, SNO+, SK+, BOREXINO, JUNO); 

- liquid (or gas) Xe detectors; 

- new ideas - !? 

Most of these approaches are used in present experiments (see above) and planned to be used in the near future experiments. If these approaches will demonstrate good results in experiments with $\sim$ 100-1000 kg of isotopes then the same approaches can be used (after some improvements) for 10 tons experiment. And, of course, new ideas can appears in the future.

\subsection{Possible background limitations}

The background conditions are the key point for the $\beta\beta$ decay experiments. To detect the 
$0\nu\beta\beta$ decay, one has to detect (as a minimum) $\sim$ 5-10 events and background must be limited to $\sim$ 0-2 events only! For HPGe detectors, the BI has to be $< 5\times10^{-6}$ c/kev$\times$kg$\times$yr ($\sim$ 200 times better than the background in the Majorana Demonstrator and GERDA-II experiments). Background can be a real problem for the next generation experiments. The main sources of background 
are the following: 

- radioactive contaminations in detector and shield; 

- cosmic rays; 

- cosmogenic activation;

- 2$\nu$ tail; 

- solar, reactor and geo neutrinos.

Of course, the necessary background level differs among experiments. But, in any case, it is 
better to have a "clever" detector, which can recognize $\beta\beta$ events (granularity, anti-coincidence, track reconstruction, daughter ion registration, etc). It is well known that in such detectors as BOREXINO, SNO and KamLAND, the purity of different liquids and gases is on the 
level $\sim$ $10^{-17} - 10^{-18}$ g/g of $^{238}$U and $^{232}$Th. One can hope that solid material can also be purified to the same level (in present experiments it is $\sim 10^{-12}-10^{-14}$ g/g)\footnote{For example, concentration of $^{238}$U and $^{232}$Th in electroformed copper produced under underground conditions is on the level of $10^{-14}$ g/g~\cite{ABG16}.}.  So, in principle, one can have pure enough materials for the 10 tons $\beta\beta$ decay experiments, but it will take a lot of effort, time and money. Let us note that the background from natural radioactivity sharply falls at energy more than 2.615 MeV. Therefore for isotopes with Q$_{\beta\beta}$ $>$ 2.615 MeV, it is easier to achieve the demanded level of a background. For the cosmic rays, 
the main background is connected with the muons themselves, $\gamma$ and neutrons induced by 
the cosmic ray muons and radioactive isotopes produced by muons. The main solution here 
is to go deep underground (6000 m w.e. or more) and to use an effective veto shield. In \cite{MEI06}, it 
was demonstrated that BI $\sim10^{-6}$ c/keV$\times$kg$\times$yr can be obtained for HPGe detectors. 
To avoid a contribution from the 2$\nu$ tail, the energy resolution has to be better than 1\% (see 
discussion in \cite{ZDE04}). It is clear that this requirement essentially narrows a circle of possible 
experiments. There are only HPGe detectors, low temperature bolometers and detectors with 
gaseous Xe. 

With a large mass of the detector's sensitive material, the background from solar, reactor, and geo-neutrinos can limit the sensitivity of the experiments. The possible contribution to the background from neutrinos was discussed in \cite{ELL04,BAR11b,BAR14,EJI14,EJI16,EJI17,EJI17a}. The main danger is connected with the solar neutrinos, since the background from the reactor and geo-neutrinos is several orders of magnitude smaller. It was shown that the contribution to the background from $\nu$e scattering is small enough and the main contribution will be given by the interaction of solar neutrinos with the nuclei of the working substance. This background depends on the specific nucleus and varies from $7.5\times 10^{-7}$ c/keV$\times$kg$\times$yr for $^{76}$Ge to $7.5\times 10^{-5}$ c/keV$\times$kg$\times$yr for $^{82}$Se. For $^{100}$Mo and $^{150}$Nd, this background will be $\sim 2\times 10^{-6}$ c/keV$\times$kg$\times$yr, and for $^{130}$Te and $^{136}$Xe $\sim 10^{-5}$ c/keV$\times$kg$\times$yr (it was considered the case of a simple calorimeter, without any additional selection of events) \cite{EJI17,EJI17a}. Using the additional capabilities of each particular detector (granularity, spatial resolution, signal selection by time correlation, track information, etc.), one can hope on additional suppression of the background by $\sim$ 10-100 times. I.e. it is hoped that for 10 tons of detector it is possible to reduce this background to a level of $\sim$ (0-1) events in the ROI region for 10 years of measurements. Nevertheless, this background presents a great challenge and must be carefully analysed when preparing new experiments and taking into account the specific features of each particular detector.


\subsection{Possibility of 10 tons isotope production}

It is 
clear that centrifugal separation is the only method which could produce 10 tons of enriched 
material for the $\beta\beta$ decay experiments in a reasonable time. Present "world" productivity can be estimated as $\sim$ 200 kg per year. One can imagine that it can be increased by $\sim$ 10 times with additional financial investment. So, 10 tons could be produced during a 5-10 year period. Possible isotopes by this moment are $^{76}$Ge, $^{82}$Se, $^{100}$Mo, $^{116}$Cd, $^{124}$Sn, $^{130}$Te and $^{136}$Xe. 

\subsection{Cost of the 10 tons experiment}
 
In Table 8 the approximate price of $\beta\beta$ isotopes is presented. Actually, the cost of the experiment includes the cost of the detector itself and this cost in the most cases can be estimated as $\sim$ 20-50 Mln. \$ (depending on the type of detector). It will include production of pure construction materials, production of the crystals in some cases, electronics, etc. From the point of view of the cost of 10 tons of isotope one can see that the cheapest possibilities (among all isotopes) are $^{136}$Xe and $^{130}$Te, $\sim$ 40-80 Mln. \$ and $\sim$ 160 Mln. \$, respectively. $^{76}$Ge and $^{100}$Mo can be used too, but they are on the border of the cost possibilities ($\sim$ 640 Mln. \$). For other isotopes, the cost starts to be a real limitation. In the case of $^{136}$Xe there is another problem. Xenon is a very rare gas, its concentration in the atmosphere is $\sim10^{-5}$ \%. The world production rate of Xe is $\sim$ 
40 tons/yr. To collect 10 tons of $^{136}$Xe one will need 100 tons of natural Xe. It means that it will be very difficult (if possible) to have 10 tons of $^{136}$Xe, but to cover the 3-5 meV sensitivity region, $\sim$ 30 tons of $^{136}$Xe will be needed (see Table 10).

\subsection{Conclusion}
Taking into account all the above arguments one can conclude the following:

1. A 10 tons detector with sensitivity to the neutrino mass at the level of $\sim$ 3-5 meV could be built 
using existing techniques. To reach 1 meV sensitivity $\sim$ 30-100 t detector (as minimum) will be needed.

2.  $^{100}$Mo, $^{116}$Cd and $^{150}$Nd are the best candidates (in the sense of sensitivity to the neutrino mass ). 10 t experiment with these isotopes will completely cover 3-5 meV region. Taking into account the price of isotopes, $^{100}$Mo remains the only acceptable candidate (low temperature bolometer detectors could be used in this case, for example).

3. From practical point of view cheapest 10 t experiment can be realised using $^{136}$Xe or $^{130}$Te. But in this case sensitivity will be $\sim$ 5 times worse in comparison with $^{100}$Mo case. To completely cover 3-5 meV region one will need $\sim$ 30-50 t of these isotopes. 
In case of $^{130}$Xe it can be EXO type detector with gaseous Xe \cite{DAN00} or NEXT \cite{FER17}/PandaX-III \cite{CHE17} type detectors. 
In case of $^{130}$Te a $^{130}$TeO$_2$ low temperature bolometer looks to be the best candidate for such experiment if existing problems with a background in such detectors will be solved (CUPID \cite{WAN15,WAN15a} approach, for example). 

In any case, one has first to investigate inverted hierarchy region (CUORE, LEGEND, nEXO, KamLAND2-Zen, SuperNEMO and other experiments).  There is a chance that $0\nu\beta\beta$ decay will be detected by these experiments. In this case main goal will be to investigate of various decay parameters in different nuclei to determine the mechanism of $0\nu\beta\beta$ decay. It will be very important to investigate the decay using tracking detectors (like SuperNEMO or gas Xe TPC) which give information about single electron energy distribution and angular distribution. 

If $0\nu\beta\beta$ decay will be not detected in inverted hierarchy region  one has to check normal hierarchy. Using present information about neutrino mixing from oscillation experiments and limit on $\sum m_i$ from astrophysical and cosmological observations in case of normal hierarchy present possible region for $\langle m_{\nu} \rangle$ is $\approx$ (1-30) meV (see, for example, \cite{SIM16}). In limiting case (mass of lightest neutrino is lower than 1 meV) $\langle m_{\nu} \rangle$ is  $\approx$ (1-5) meV.  In this case it will be necessary to investigate $\sim$ (10-100) tons of isotope. From the standpoint of today only experiment with 10 tons of isotope looks possible. It is impossible to produce 30-100 tons of isotope in a reasonable time and with reasonable cost using existing technique. So, to do this, some radical improvements in isotope production technique are needed. 

But if mass of lightest neutrino is high enough (say, 5-26 meV) there is a chance that for normal hierarchy case $\langle m_{\nu} \rangle$  is on the level $\sim$ (8-30) meV. In this case it will be much more easy to detect the $0\nu\beta\beta$ decay and 1-10 tons detector can see the effect.

\section{References}
\hspace{0.5cm}

\end{document}